\documentstyle[12pt,aasms4]{article}

\begin{document}

\def\pdot {\dot P}
\def\Omdot {\dot \Omega}
\def\ltsima{$\; \buildrel < \over \sim \;$}
\def\lsim{\lower.5ex\hbox{\ltsima}}
\def\gtsima{$\; \buildrel > \over \sim \;$}
\def\gsim{\lower.5ex\hbox{\gtsima}}
\def\msole{~M_{\odot}}
\def\mdot {\dot M}
\def\axj {AX~J0851.9$-$4617.4~}
\def\rxj {RX~J0852.0$-$4622~}
\def\sax  {SAX~J0852.0$-$4615~}

\title{The X--ray Sources at the Center of the Supernova Remnant \rxj}
\author{S. Mereghetti}
\affil{Istituto di Fisica Cosmica G.Occhialini - CNR \\
v.Bassini 15, I-20133 Milano, Italy}
\authoremail{sandro@ifctr.mi.cnr.it}

\begin{abstract}
We present a BeppoSAX observation of the X--ray source \axj that has
been proposed as the neutron star associated  to the shell like-supernova
remnant \rxj.
The image  at E$>$5 keV  shows the presence of a new source, \sax,
not resolved from  \axj in previous observations.
The improved positional accuracy obtained for \axj confirms its possible
identification with an early type star, also consistent with its
  soft thermal spectrum (kT$\sim$1.3 keV).
\sax, with a  harder X-ray spectrum and  a higher X--ray to optical flux ratio,
is a more likely candidate for the neutron star associated to the supernova remnant.

\end{abstract}


\section{Introduction}

The supernova remnant \rxj  consists of a circular X--ray shell, with an angular diameter
of $\sim$2$^{\circ}$, seen in projection against the much more extended
soft X--ray emission of the Vela supernova remnant (Aschenbach 1998).
\rxj (also known as G266.2--1.2) has  attracted considerable interest since Iyudin et al. (1998)
reported its possible detection
in the $\gamma$--ray  line   of $^{44}$Ti (1.156 MeV).
This isotope,   produced in different types of supernovae, has a
lifetime of $\sim$90 years and should therefore be a good tracer for
young supernova remnants (SNRs).
The observed $\gamma$--ray line flux and the relatively small angular dimensions
of the remnant imply an age of only $\sim$680 years and a
small distance of $\sim$200 pc (Aschenbach, Iyudin \& Sch\"{o}nfelder 1999).
Thus \rxj could be the remnant of the closest supernova event to have occurred in
recent historical times.

However, this interpretation poses   a few problems. For instance,
the lack of records for an event that could have reached the  optical brightness
of the full moon (for a Type Ia SN), and in any case was brighter than Venus
(for an extremely sub-luminous SN), is puzzling.
If a neutron star was formed, it should still be a bright source of thermal
emission, but no  bright soft X--ray sources have been found within \rxj
in the ROSAT All Sky Survey data. The upper limits reported by Aschenbach et al. (1999)
imply surface temperatures smaller than 3$\times$10$^5$ K,
assuming the low interstellar absorption expected for
a  distance of only a few hundreds parsecs.

Recent   observations of \rxj obtained with the ASCA satellite
have been reported by Slane et al. (2001). These new data
clearly indicate the non-thermal nature of the X--rays from the
SNR shell.  The high absorption  suggests that \rxj is
at a distance of the order of $\sim$1-2 kpc, well
beyond the Vela SNR. Consequently an age of several thousands year
is more likely.
These conclusions are also supported by the fact that
a new analysis of the COMPTEL data (Sch\"{o}nfelder et al. 2000) has reassessed the
statistical significance of the $^{44}$Ti emission
from the direction of \rxj. Depending on the data selection and method of analysis,
the excess is present at a level between $\sim$2 and 4
$\sigma$: if \rxj is not emitting $^{44}$Ti $\gamma$-rays there are no compelling reasons
to require a particularly small distance and age.

The ASCA observations revealed also a central
point source, surrounded by diffuse X--ray emission, that might be
the  neutron star associated to \rxj.
Here we report on a recent observation pointed
toward the central region   of \rxj performed with the BeppoSAX X--ray satellite.

\section{Observations}

The BeppoSAX observation was done on 1999 May 18-20.
Standard screening of the data resulted in a net exposure time of $\sim$78 ks
in the Medium Energy Concentrator Spectrometer (MECS) instrument.
The MECS (Boella et al. 1997), consisting
of two nearly identical imaging gas scintillation  proportional counters coupled to
grazing incidence X--ray telescopes, operates in the 1.8-10 keV energy range.

A relatively bright source
($\sim$0.017 counts s$^{-1}$)
is detected near the center of the MECS field of view.
Its position  is
RA: 8$^h$ 52$^m$ 01$^s$, Dec: --46$^{\circ}$ 18$'$ 14$''$ (J2000) with an
uncertainty of 1$'$.
The source is very soft and is only visible at energies smaller than $\sim$5 keV (Fig. 1a).
The hard X--ray image (5-10 keV, Fig. 1b)
shows instead   the presence at RA: 8$^h$ 52$^m$ 01$^s$, Dec: --46$^{\circ}$ 14$'$ 58$''$
of a second, weaker source that we designate \sax.
To evaluate the statistical significance of this source we have used
the detection algorithm implemented in the XIMAGE V3.01 software package.
At E$>$ 5 keV, \sax is detected at a level of $\sim$7$\sigma$ above the local
background.
The angular separation between the two sources is more than 3$'$,
significantly greater than the statistical error associated to the source
positions. The different location of the two sources  is also illustrated in Fig. 1a,
where the contour levels of the hard
band (5-10 keV) are overlaid on the soft X--ray image.

The position of the bright, softer source is consistent  with that of
the excess seen in the ROSAT soft X--ray ($<$2.4 keV) data (Aschenbach 1998) and with
that of the source detected with ASCA (Slane et al. 2001).
In the following we will refer to this source by its ASCA name, \axj.
The limited angular resolution and shorter exposure time of the ASCA
observation of \rxj  did not allow   \sax to be detected as a separate source.
The second weak soft X--ray (0.1-2.4 keV) source   reported by
Aschenbach et al. (1999)   lies more than 3$'$   to
the south of the sources discussed here and is not visible in the MECS images.

The BeppoSAX satellite carries also the LECS instrument that is similar to a
single MECS unit, but covers a lower energy range (0.1-10 keV).
We do not see evidence for point sources in  the LECS image
(exposure time $\sim$28 ks), which
below 2 keV is dominated by the diffuse emission from the Vela SNR.

Using the MECS data we performed a spectral analysis of \axj in
the 1.8-10 keV range.
The source counts were extracted from a circular region with radius of 4$'$ and
rebinned in order to have at least 30 counts for each spectral channel.
The background spectrum was estimated from a region surrounding the source,
since the standard MECS background spectra, obtained from
empty fields at high galactic latitude, are not appropriate in this
region dominated by the diffuse emission
from the Vela SNR.

Equally good fits could be obtained with a power law (photon index $\alpha$ = 3.6$\pm$0.6),
with a thermal bremsstrahlung (kT$_{Br}$ = 1.3$\pm$0.4 keV), and with a blackbody
(kT$_{BB}$ = 0.5 $\pm$ 0.1 keV).
The absorption value is not well constrained by the data:
only   upper limits of a few $\times$10$^{22}$ cm$^{-2}$ could be derived (see Fig. 2).
The unabsorbed 2-10 keV flux for the best fit power law
parameters is 6.6$\times$10$^{-13}$ erg cm$^{-2}$ s$^{-1}$.
The corresponding values for the bremsstrahlung and blackbody models
are 6.0 and 5.6 $\times$10$^{-13}$  erg cm$^{-2}$ s$^{-1}$,
respectively.
We verified that the inclusion of the LECS data in the fit did not
improve the determination of the spectral properties.

The light curve of \axj does not show evidence for variability
during the BeppoSAX observation. We performed also a search for periodicity
in the range from 0.01 to 500 s with negative results.
Comparing our results with those obtained during the ASCA
observation (Slane et al. 2001), we find no evidence for long term flux variations.
In fact the MECS values of $\alpha$, kT$_{BB}$ and absorption are consistent with
the ASCA ones.
Fixing the parameters to the ASCA best fit values  ($\alpha$=3.2 and
N$_H$=3.7$\times$10$^{21}$ cm$^{-2}$), we derive a
0.5-10 keV  flux of 2$\times$10$^{-12}$ erg cm$^{-2}$ s$^{-1}$
(corresponding to 5$\times$10$^{-12}$ erg cm$^{-2}$ s$^{-1}$, corrected for the absorption).
This is, within the uncertainties, compatible with that measured with ASCA five months earlier.

The second source, \sax, with only $\sim$80 net counts detected
above 5 keV, is too weak for a detailed  spectral and timing analysis.
It is clearly harder, and possibly more absorbed than  \axj.
Assuming a power law spectrum with photon index $\alpha$ between 1 and 3 and
N$_H$=4$\times$10$^{21}$ cm$^{-2}$
(i.e. the absorption derived from the fits to the \rxj shell
(Slane et al. 2001)),  the observed flux in
the 2-10 keV range lies between
$\sim$ 3 and 4 $\times$10$^{-13}$ erg cm$^{-2}$ s$^{-1}$.

The BeppoSAX error regions of the two sources are
shown by the small circles (1$'$ radius) overlaid in Fig. 3
on an optical image obtained from the UK Schmidt Digitized Sky Survey.
The new position of \axj is consistent with that previously
derived with ASCA, which had an uncertainty of 2$'$ (larger circle in Fig. 3).
As noted by Aschenbach (1998) and Slane et al. (2001),
two early type stars are consistent with the position of \axj.
Both stars, HD76060 and Wray 16-30,  lie inside the smaller
error circle obtained with BeppoSAX. The brightest object in the
error circle of \sax has a magnitude $\sim$15.

Slane et al.  (2001) noted that \axj is surrounded by diffuse X--ray emission
with a non-thermal spectrum.  This emission is also clearly detected in our
BeppoSAX observation (see Fig.~1).  To study its spectrum we extracted the MECS
2-10 keV counts from the inner part of the detector (radius 8$'$) excluding
circular regions with 2$'$ radius centered on the positions of \sax and \axj.
The background spectrum was estimated from two regions in the outer part of the
MECS field of view.  
After a rebinning to obtain at least 30 counts per spectral
bin, the data were fitted using the response matrix and effective area values
appropriate for this complex extraction region.  
The best fit was obtained with a power
law ($\alpha$=1.8 $\pm$ 0.5, N$_H$ $<$ 2$\times$10$^{22}$ cm$^{-2}$, 
F$_{2-10 keV}$ = 1.5$\times$10$^{-12}$ erg cm$^{-2}$ s$^{-1}$ unabsorbed).  
For $\alpha$=2 and
N$_H$=1.15$\times$10$^{22}$ cm$^{-2}$, the unabsorbed flux in the 0.5-10 keV
range is F$_{0.5-10 keV}$ $\sim$3$\times$10$^{-12}$ erg cm$^{-2}$ s$^{-1}$.
The difference with respect to the corresponding value reported by Slane et al.
(6.7$\times$10$^{-12}$ erg cm$^{-2}$ s$^{-1}$) cannot be due to \sax
(if we do not exclude its region from the fit of the diffuse emission
we obtain F$_{0.5-10 keV}$ $\sim$3.7$\times$10$^{-12}$ erg cm$^{-2}$ s$^{-1}$ ).
The discrepancy is probably
to be ascribed to a different evaluation of the (highly uncertain) background
and of the contribution of the central source(s), as well as to uncertainties
in the relative calibration of the two instruments for what concerns
the response to regions of diffuse emission.
In any case, independent on the absolute luminosity of the diffuse emission,
it is important to note that our results confirm its non-thermal origin.

\section{Discussion}

Both \axj and \sax, located close to the geometrical center of
the SNR,  are in principle good candidates to be the neutron star 
associated to \rxj.
However, as shown in the following   discussion,
the interpretation of either source as a neutron star is
problematic in two respects.
First, since none of them has obvious optical counterparts, 
even the source not interpreted as a neutron star must have some 
peculiarity. 
Second, independent of which of the two sources is associated
to the SNR, very unusual neutron star properties are required if   \rxj
is really as young and nearby ($\sim$680 years, d$\sim$200 pc) as
suggested by Iyudin et al. (1998).

\subsection{\axj}

Both Aschenbach  (1998) and Slane et al. (2001) considered whether
one of the two early type stars could be the counterpart of \axj.
Though they concluded that a neutron star interpretation is more likely,
the fact that the new smaller error circle of \axj still contains HD76060
and Wray 16-30, leads us to briefly discuss again these stars as
possible counterparts.

HD76060, is a bright  (V=7.88, B--V= --0.09)
B8 type star of luminosity class IV-V.
Based on  the analysis of its parallax (3.7$\pm$0.66 mas)
and proper motion, de Zeeuw et al. (1999) concluded that HD76060
belongs to the OB association
Trumpler 10  that is at
an average distance of $\sim$370 pc.
Late B stars of luminosity class III-V can reach values of
L$_x$/L$_{bol}$ as high as 10$^{-5}$, possibly higher in the case
of X--ray selected objects as in the case discussed here.
Thus HD76060 could contribute to  part
of the observed X--ray flux,  but it is unlikely that this star is
responsible for the whole luminosity ascribed to \axj.

A more promising candidate is Wray 16-30,
a  B[e] star with V=13.8  (Th$\acute{e}$, de Winter \& P$\acute{e}$rez 1994)
also detected with IRAS
(IRAS 08502--4606).
The 0.5-10 keV luminosity of \axj is   between 1
and 4 $\times$ 10$^{32}$ d$_{kpc}^2$ erg s$^{-1}$,
where the large uncertainty depends on
the assumed amount of interstellar absorption
(d$_{kpc}$ is the distance in units of kiloparsecs).
Since the  apparent magnitude of Wray 16-30 suggests a distance
in excess of 1 kpc,
the corresponding X-ray luminosity is relatively high,
compared to the typical values   observed in Be stars
(Damiani et al. 1994, Zinnecker \& Preibisch 1994).
Note however that such luminosity values (10$^{29}$ -- 10$^{32}$  erg s$^{-1}$)
were obtained from   observations of Ae/Be stars in the soft X--ray range.
Therefore they might be biased toward low values,
due to the substantial effect of absorption in the circumstellar dust
halos often  surrounding these stars, as testified by the presence
of IR excesses in their spectra.
Thus the association of \axj with Wray 16-30 cannot be completely ruled out.

It is also possible that  Wray 16-30 is in a binary system
with an accreting neutron star.
Be/neutron star systems  are the most
common X--ray binaries in the Galaxy (see, e.g., Coe 1999).
During outburst they  can reach a luminosity
of 10$^{36}$ - 10$^{37}$  erg s$^{-1}$, but most of the time they emit at
a quiescent level typically in the range 10$^{33}$ - 10$^{35}$  erg s$^{-1}$
(Neguerela 1998).
If  Wray 16-30 has a compact companion,  the   X--ray flux of \axj could
be easily explained even for d$_{kpc}>$ 1 kpc.
The chance probability of finding a Be/neutron star system within a few
arcmin from the center of \rxj is difficult to estimate. However, it is probably
low enough to suggest an association with the SNR.

In the alternative interpretation of  \axj as an isolated neutron star produced
in the  \rxj supernova explosion, 
interesting constraints can be derived (see also Slane et al.  2001).

If \rxj is indeed as young as implied by  
the $^{44}$Ti results, we could expect a  young, energetic
neutron star,  emitting mostly by non-thermal processes of
magnetospheric origin.
Assuming the small distance (d$\sim$200 pc) 
proposed by Aschenbach et al. (1999), 
the X-ray luminosity would be extremely small
for a young pulsar ($\sim$3 10$^{30}$  erg s$^{-1}$).
A more reasonable luminosity value is instead obtained if \axj
is at the distance of $\sim$1-2 kpc  derived by Slane et al. (2001)
for \rxj .
In any case,  
the best fit power law index $\alpha$=3.6 is much steeper than that observed
in objects of this type, like, e.g., the Crab pulsar or PSR 0540--69
(see, e.g., Becker \& Tr\"{u}mper 1997).

A different possibility is that the X--rays are of thermal origin.  In this case
we can use the results of the blackbody fit to constrain the distance: d$\sim$7
R$_{km}$ kpc, where R$_{km}$ is the radius of the emitting region in km.
It is clear that emission due to   thermal cooling
from (a large fraction of) the whole neutron star surface  
can be excluded since it would lead to an unplausible distance.
Futhermore, the observed blackbody  
temperature, T$_{BB}$$\sim$6 10$^6$ K, is extremely
high  even for a young pulsar (see, e.g., Page 1998).
This  problem could be partly reduced, but   not completely
solved, using more realistic
models for the neutron star atmosphere thermal emission, that typically yield
lower effective temperatures and smaller distances than blackbody fits
(see, e.g., Zaviln, Pavlov \& Tr\"{u}mper 1998).
 
Thermal components with high blackbody temperatures 
(T$_{BB}$$\sim$2-4$\times$10$^6$ K) have been 
observed in old, nearby radio pulsars such as  PSR~B0950+08 and PSR~B1929+10
(Manning \& Willmore 1994; Yancopoulos, Hamilton  \& Helfand 1994).
They are interpreted in terms of small polar caps,
with radius of the order of a few tens of meters, heated by high-energy
particles accelerated in the pulsar magnetosphere.
A similar possibility cannot be ruled out for \axj 
(it would be at 200 pc for a polar cap radius of 30 m).
Note however that PSR~B0950+08 and PSR~B1929+10 have   characteristic ages
much larger than the typical lifetime of any SNR. 

In conclusion, any interpretation of \axj as an
isolated neutron star 
leads to some puzzle. Such problems
are probably due to the limited quality of the current data in terms of
spectral and spatial resolution, and could 
probably be  solved by a decomposition
of the observed flux into a combination of thermal and non-thermal processes,
as observed in other better studied neutron stars.

\subsection{\sax}

The brightest optical object in the error box of \sax has a magnitude $\sim$15, implying an
X--ray to optical flux ratio $\gtrsim$0.1, that makes a stellar identification unlikely
(see, e.g. Maccacaro  et al. 1988).

Though the poor statistics prevents a spectral analysis of \sax, it is clear that
this source is harder than \axj.
The interpretation in terms of a young, thermally emitting isolated neutron star
(with a temperature necessarily higher than that of \axj) leads to  the
same problems that have  been  discussed above for \axj .

The detection of X--rays with energy greater than 5 keV is more suggestive
of non thermal processes.
For a typical value of $\alpha$=2 and N$_H$=4$\times$10$^{21}$ cm$^{-2}$
(the absorption derived for \rxj)
the luminosity at the source in the 0.1-10 keV range
is   $\sim$10$^{32}$ d$_{kpc}^2$ erg s$^{-1}$,
consistent with a Vela-like pulsar (age $\sim$ 10$^4$ -- 10$^5$ years)
at a distance of a few kpc (see, e.g. Becker \& Tr\"{u}mper 1997).
In this case the association with \rxj  would strongly argue
against the scenario of a very young and nearby SNR (Aschenbach et al. 1999),
while it would be consistent, considering the uncertainties, with
the larger distance and age inferred from the ASCA observations of
\rxj (Slane et al. 2001). 

The diffuse X--ray emission, with a  power law spectrum, 
can be interpreted as a
synchrotron nebula surrounding the central neutron star. 
It  extends to the north of \axj and is roughly centered
around \sax. A detailed study of its morphology is complicated by the presence
of the soft diffuse emission associated to the Vela SNR, and by the poor angular and
spectral resolution of these data. Future studies with Chandra and NewtonXMM will be
crucial in this respect.

Finally we note that \sax is located
$\sim$7$'$ from the     geometrical center of \rxj
(RA: 8$^h$ 52$^m$ 03$^s$, Dec: --46$^{\circ}$ 22$'$,   Aschenbach (1998)).
The angular displacement from the site of the explosion can be
easily explained with a reasonable value for the
neutron star transverse
velocity of $\sim$200 (D$_{kpc}$/$t_{10 kyr}$) km s$^{-1}$.

\section{Conclusions}

A recent BeppoSAX observation of  the central region of the
supernova remnant \rxj has led to the detection of two X--ray sources that can
be considered good neutron star candidates.
The softer source, \axj was already observed with ROSAT and ASCA.
Though unlikely (Slane et al. 2001), it cannot be excluded that the X--rays
from this source are due
to one (or both) of the two early type stars present in its error region.
The newly discovered X--ray source, \sax,
does not have obvious stellar counterparts.
It  has a harder spectrum and
properties  consistent
with a  neutron star with an age of a few 10$^4$ years at a distance of a few kpc.

Unless very unusual properties are invoked,
both sources are difficult to explain as the neutron star associated to \rxj, if the
latter is really as young and nearby ($\sim$680 years, d$\sim$200 pc) as
implied by the claimed detection of a $^{44}$Ti  $\gamma$--ray line
(Iyudin et al. 1998).

\acknowledgments

This work was based on data obtained through the BeppoSAX Data Center provided
by the Italian Space Agency. The referee Patrick Slane provided useful
comments that helped to improve this paper.

\clearpage

\figcaption[sgi9259.eps]
{BeppoSAX MECS images in the 2-5 keV (left) and 5-10 keV (right)
of the sources at the center of \rxj. Both images have been smoothed with a
Gaussian with $\sigma$=1$'$. The highest contour level of the 5-10 keV image is
drawn  on the soft X--ray image to show the different positions of the two
detected sources.
\label{fig1}}

\figcaption{Confidence contours (68\%, 90\% and 99\%) for the spectral parameters of \axj derived
from the MECS fits; (a) power law, (b) thermal bremsstrahlung, (c) blackbody.
\label{fig2}}

\figcaption[sgi9279.eps]{Error regions of the sources near the center of \rxj
plotted on the digitized Sky Survey image. North is to the top and East to the left.
The BeppoSAX error circles have a 1$'$ radius. The northern one is for the newly
discovered source \sax.
The ASCA error circle for \axj has a 2$'$ radius.
\label{fig3}}


\plotone{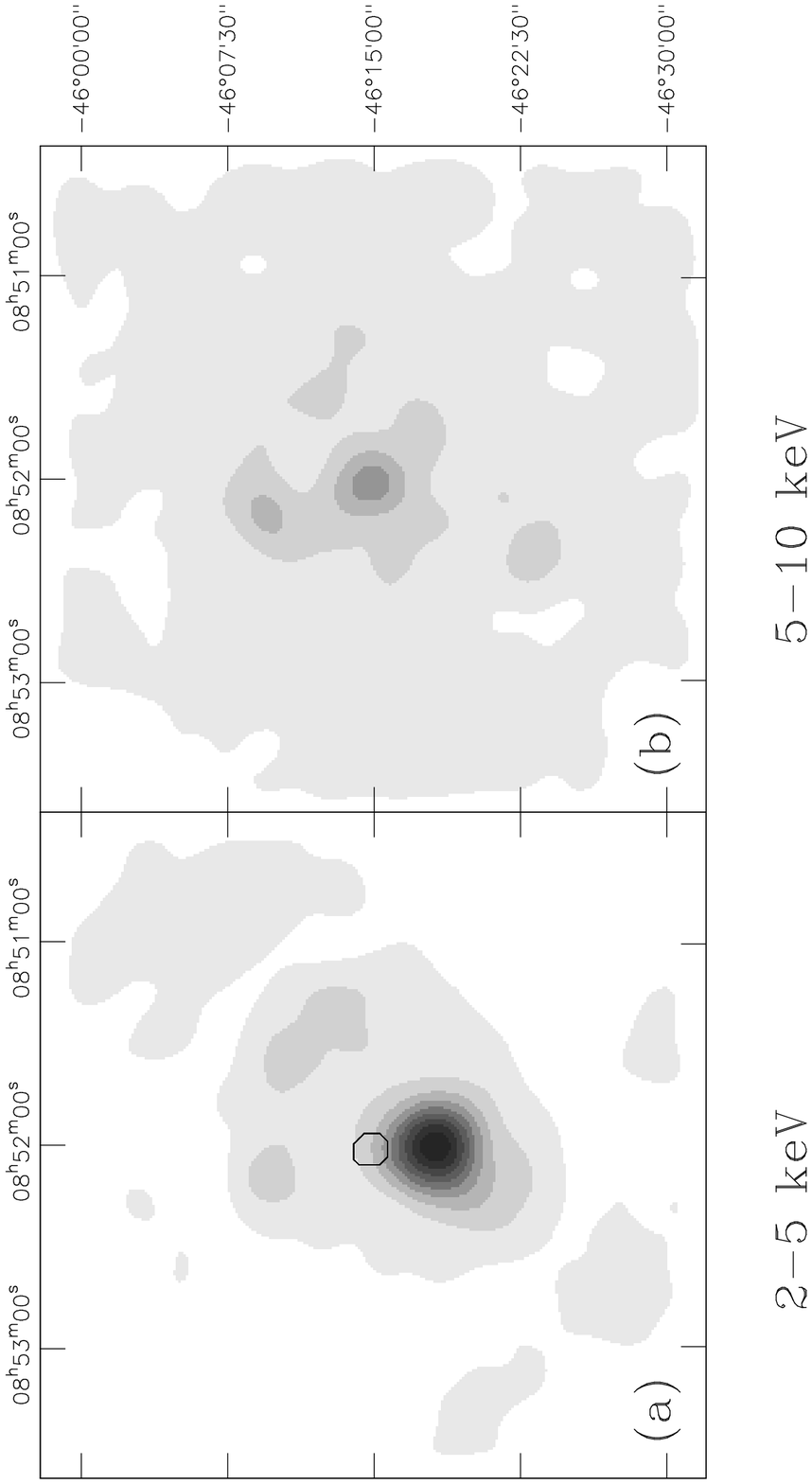}


\plotone{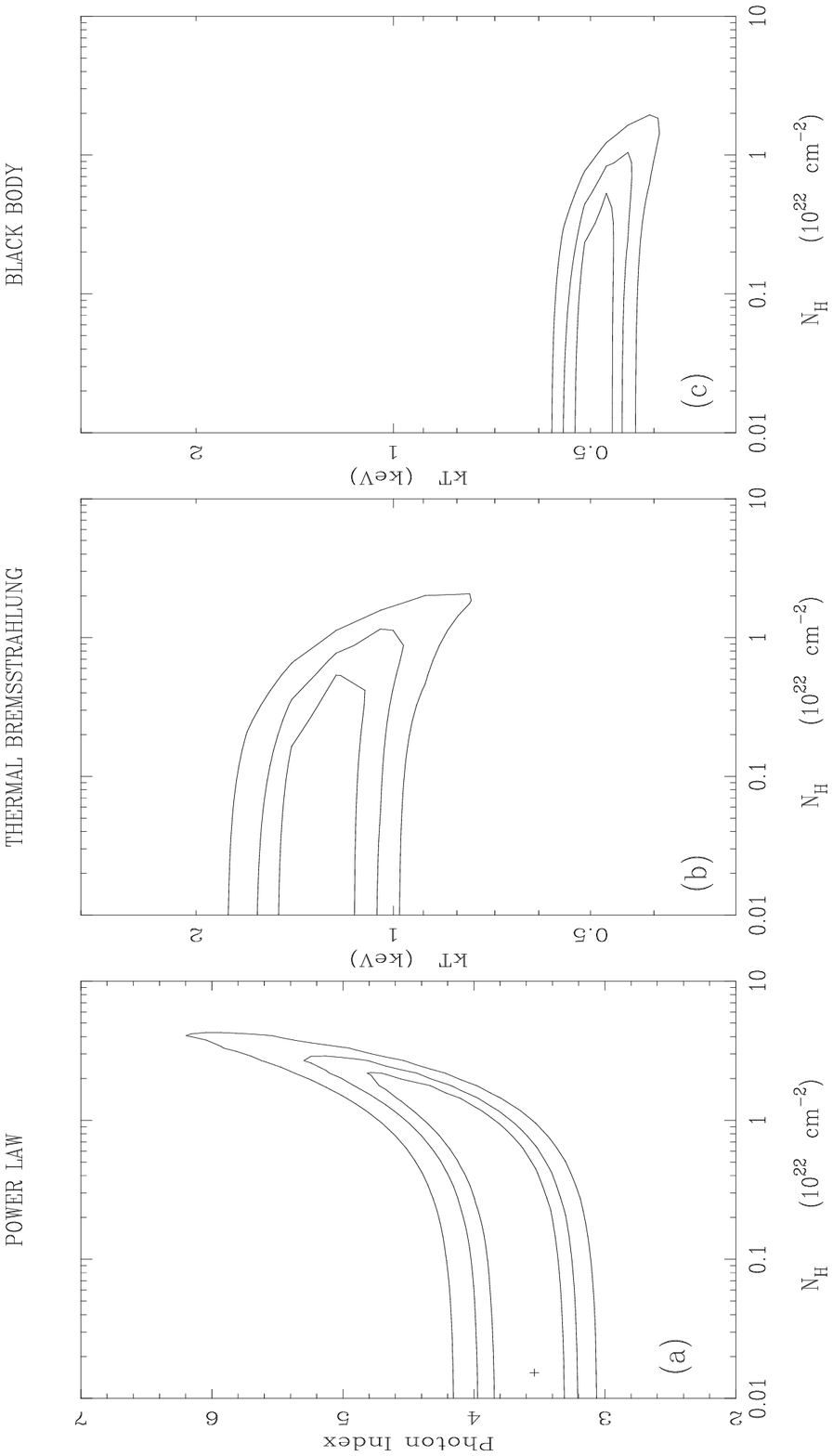}

\clearpage

\plotone{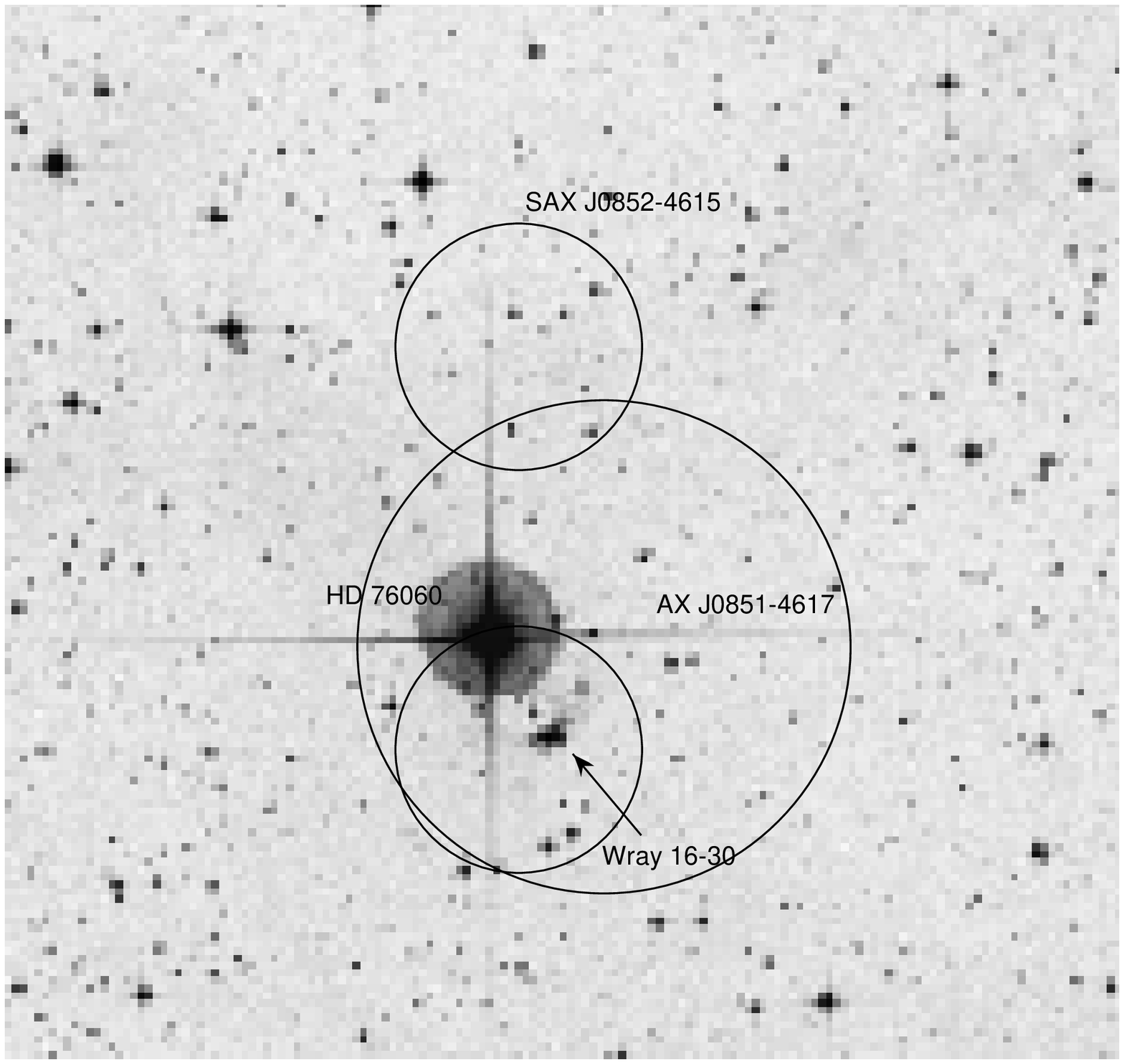}


\begin{references}
\reference{} Aschenbach, B.  1988, \nat, 396, 141
\reference{} Aschenbach, B., Iyudin A.F., \& Sch\"{o}nfelder V. 1999, A\&A, 350, 997
\reference{} Becker W. \& Tr\"{u}mper J. 1997, A\&A, 326, 682
\reference{} Boella G. et al. 1997, A\&AS, 122, 299
\reference{} Coe M.J. 1999, astro-ph/9911272
\reference{} Damiani F., Micela G., Sciortino S., \& Harnden F.R.Jr. 1994, \apj, 436, 807
\reference{} de Zeeuw P.T. et al. 1999, \aj,  117, 354
\reference{} Iyudin A.F. et al. 1998, \nat, 396, 142
\reference{} Maccacaro T. et al. 1988, \apj, 326, 680
\reference{} Manning R.A. \& Willmore A.P. 1994, \mnras, 266, 635 
\reference{} Negueruela I. 1998, A\&A, 338, 505
\reference{} Page D. 1998, in \textit{The Many Faces of Neutron Stars},
R. Buccheri, J. van Paradijs, \& M. A. Alpar eds.,   Kluwer Academic Publishers,   p.539
\reference{} Sch\"{o}nfelder V. et al. 2000, Proc. 5$^{th}$ Compton Symposium,
M.L.McConnell \& J.M.Ryan eds., AIP Conf. Proc. 510, 54
\reference{} Slane P. et al. 2001, \apj, in press, astro-ph/0010510
\reference{} Th$\acute{e}$ P.S., de Winter D. \& P$\acute{e}$rez M.R. 1994, A\&ASS, 104, 315
\reference{} Yancopoulos S., Hamilton T.T., \& Helfand D.J. 1994, \apj, 429, 832
\reference{} Zaviln V.E., Pavlov G.G. \& Tr\"{u}mper J. 1998, A\&A, 331, 821
\reference{} Zinnecker H. \& Preibisch T. 1994, A\&A, 292, 152
\end{references}
\end{document}